\documentclass[twocolumn,pra,showpacs,amsmath]{revtex4}
\usepackage{graphicx}
\newcommand{\ua}{\uparrow}

\DeclareMathOperator{\Acosh}{cosh^{-1}}
\begin{document}

\author{Jonathan Keeling}
\affiliation{Cavendish Laboratory, University of Cambridge,
  J.~J.~Thomson Ave., Cambridge CB3 0HE, UK}
\email{jmjk2@cam.ac.uk}
\title{Quantum corrections to the semiclassical collective dynamics in the Tavis-Cummings model}
\begin{abstract}
The Tavis-Cummings model (the Dicke model treated in the rotating wave
approximation) describing many two-level systems coupled to a single
bosonic mode, has been long known to show collective semiclassical
oscillations when prepared in an inverted state, with all two-level
systems excited, and the bosonic mode empty.  This paper discusses how
the quantum dynamics approaches this semiclassical result for large
numbers of two-level systems, focussing on how the eigenvalues
approach their semiclassical limit.  The approach to the semiclassical
result is found to be slow, scaling like a power of the logarithm of
the system size.  Considering also the effect of weak detuning between
the two-level system and the bosonic field, quantum corrections are
again found to decay slowly with system size, such that for a fixed
detuning, the quantum effects of detuning are greater than the
classical effect.
\end{abstract}
\pacs{%
42.50.Nn, 
42.50.Pq, 
03.75.Kk, 
03.75.Lm  
}
\maketitle

\section{Introduction}
\label{sec:introduction}

The Dicke model~\cite{dicke54}, describing interaction between a
number of two-level systems and a single bosonic mode, has long been
studied as a simple model of cavity quantum electrodynamics, which
despite its simplicity can show quite intricate behaviour.  When the
coupling between the two-level system and the bosonic mode is treated
in the rotating wave approximation, the Dicke model reduces to the the
Tavis-Cummings model~\cite{tavis68}\footnote{In the context of
  superradiance and of atom-molecule interconversion in cold atoms,
  the Tavis-Cummings model is often referred to as the Dicke
  model.}. One long studied feature of this model is collective
oscillations that arise when the initial state is inverted; the
simplest such case concerns an initial state with an empty bosonic
mode, and all two level systems in their excited state.  In a
semiclassical approximation~\cite{bonifacio70} the number of bosons
describes a train of hyperbolic secant pulses.  Interest in these
collective oscillations has recently been revived both by the
connection to atom-molecule interconversion in cold atomic
gases~\cite{andreev04,barankov04,yuzbashyan05:prb,szymanska05:bec-bcs},
as well as potential experiments studying coupling between quantum
dots and a cavity photon modes~\cite{eastham07,eastham08}.  The
possibility of coupling between a radiation mode and multiple
two-level systems is also being pursued in Circuit-QED
experiments~\cite{fink09}, in which the two-level systems are
superconducting qubits.  Both these latter examples are closely
related to the original context of this problem~\cite{bonifacio70}:
two-level atoms coupled to a photon mode in a cavity.  This problem is
also closely related to collective superfluorescence~\cite{gross82}
for initially inverted atoms but without a cavity.  Without the cavity
there is only a single hyberbolic secant pulse since the dense
spectrum of photon modes prevents recurrence.


The aim of this article is to study the quantum dynamics of the Tavis-Cummings
model with a finite number of two level systems $N$, starting from a
fully inverted state, in order to see how the quantum dynamics differ
from the semiclassical dynamics.  In particular, considering the case
of a completely symmetric Tavis-Cummings model (where all two-level systems are
identical), one finds that the semiclassical results are recovered in
the limit of an infinite number of two-level systems, but that the
approach to this semiclassical limit is slow, scaling as a power of
the logarithm of the number of two-level systems. 
In addition to the sequence of hyberbolic secant pulses that exist in
the semiclassical dynamics, the quantum dynamics is found to have an
additional slow envelope.  As the number of two-level systems
increases, the period of this envelope increases compared to the
period of the train of hyberbolic secant pulses and so its effects
become negligible, however this trend is very slow, with
$T_{\text{envelope}}/T_{\text{pulse}} \sim [\ln(\sqrt{N})]^3$. 

A closely related question has been addressed by \citet{faribault},
for the case of the Richardson model rather than the Tavis-Cummings
model.  Their work focussed on how the integrability of the quantum
model allows one to calculate overlaps between the initial state and
the eigenstates, as well as matrix elements of the physical
observables.  The current work addresses a complementary question, that
of how the eigenvalues approach the semiclassical results for large
system sizes.   

Aspects of the behaviour of the Dicke model in the absence of the
rotating wave approximation have also been studied; in particular,
features of the finite size system associated with the quantum phase
transition in the infinite system size model have been considered.
These include: changes of statistics of excited state
energies~\cite{emary03:prl,emary03:pre}; perturbative approaches in
the limit of large coupling strengths~\cite{frasca04}; entanglement
between the bosonic mode and the two-level systems~\cite{lambert04},
including how that entanglement scales with the system size; and the
scaling of other quantities, such as ground state energies or
excitation gap with system size~\cite{vidal06,liu08}.  The results of
these previous studies differ from the question addressed in the
current work for two reasons: firstly, without the rotating wave
approximation, the number of excitations in the system is no longer
conserved, and so the classical problem is no longer integrable.
Secondly, the results in this article relate to the collection of
eigenstates near $|E_q|=0$, while the ground state, or the
thermodynamics at low temperatures~\cite{hepp73}, involves eigenstates
with much lower energy.

An alternate method to include quantum corrections to the
semiclassical dynamics is by accounting for the dynamics of higher
cumulants (as well as expectations) of the collective operators has
been discussed by Vardi et al.~\cite{vardi01,vardi01:pra} in related
but different models.  A similar idea has also been discussed in the
context the BCS model~\cite{szymanska05:bec-bcs,matyjaskiewicz}, but
in cases where the semiclassical dynamics is more complicated. Another
related problem concerns quantum dynamics in the central spin model
--- where a large nuclear spin (analogous to the bosonic field here)
is coupled to a sea of electronic spins (analogous to two-level
systems); Ref.~\cite{tsyplyatyev08} considers the quantum dynamics of
this model with initial conditions such that there is only a single
excitation in the system.  Other related work concerns the dynamics of
the Tavis-Cummings model in the opposite limit, of a small number of
two-level systems, starting from an initial coherent boson state,
studied in Refs.~\cite{seke89,ramon98,chumakov96} in connection to the
collapse and revival of Rabi oscillations in the case of a single
two-level system.

The results presented in this paper refer mainly to the completely
symmetric Tavis-Cummings model, for which all two level systems are identical.
This restriction allows one to extract simple analytical formulae for the
scaling of quantum corrections with system size.  In addition, this
symmetric case can be expected to be the ``most classical'' limit of
the Tavis-Cummings model, as one may derive the semiclassical dynamics in such
a case by replacing large quantum spins by classical spins. The fact
that quantum corrections exist in this most classical case suggest that
important corrections may exist in the non-symmetric Tavis-Cummings model.  One
specialised limit of this is discussed in
Sec.~\ref{sec:rela-detun-disord}, supporting this idea.

The rest of the paper is organised as follows;
section~\ref{sec:hamilt-numer-results} introduces the Hamiltonian, and
discusses the previously known results of the semiclassical
approximation; these are compared to the results of exact
diagonalisation in Fig.~\ref{fig:simple-trace}.
Section~\ref{sec:wkb-appr-scal} then shows how the quantum corrections
can be extracted from a WKB approach to the problem, focussing on the
case where the two-level system and boson energies match; the effect
of detuning is discussed in section~\ref{sec:detuning}, and concluding
remarks are given in section~\ref{sec:conclusions}.

\section{The model and comparison of semiclassical and numerical results}
\label{sec:hamilt-numer-results}


The model studied in this paper can be written:
\begin{equation}
  \label{eq:dicke}
  H =
  \sum_{i=1}^N
  \left(
    \epsilon_i s_i^z + 
    s_i^+ a^{} + s_i^- a^{\dagger} 
  \right), 
\end{equation}
where the spin operators obey $[s_i^z, s_i^{\pm}] = \pm s_i^{\pm},
[s_i^+,s_i^-]=2s_i^z$ and $a,a^\dagger$ are bosonic operators.  The
coupling between the two-level systems and the bosonic mode has been
scaled to $1$, hence all other energies and times are measured in
units of this coupling.  The initial state of the system is taken to
be $|n=0, \ua\ua\ua\ldots\rangle$, where all the two-level systems are
excited, and the bosonic mode is empty.  In the subsequent dynamics
there are collective oscillations, transferring excitations between
the two-level systems and the bosonic mode.

For comparison to the exact dynamics, the following briefly summarises
the semiclassical solution, described in
Refs.~\cite{andreev04,barankov04}.  The semiclassical equations
correspond to writing the Heisenberg equations of motion for the
operators $s_i^-, s_i^z, a$, and then replacing these operators by
commuting classical variables.  The resultant equations can be solved
by the ansatz:
\begin{equation}
  \label{eq:2}
   s_i^- = 
  \frac{(\epsilon_i - \mu) a + i \dot{a}}{%
    (\epsilon_i - \mu)^2 + \lambda},
  \quad
   s_i^z  = \frac{1}{2} 
  - \frac{a^2}{(\epsilon_i - \mu)^2 + \lambda}
\end{equation}
along with the equation of motion for $a$, $\dot{a}^2 = a^2(\lambda -
a^2)$, and the self consistency conditions for $\mu, \lambda$:
\begin{equation}
  \label{eq:1}
  1= \sum_i \frac{1}{(\epsilon_i - \mu)^2 + \lambda},
  \quad
  \mu = \sum_i \frac{\epsilon_i - \mu}{(\epsilon_i - \mu)^2 + \lambda}.
\end{equation}
These equations are analogous to the BCS gap equation, with $\mu$
acting as a generalised chemical potential (i.e.\ common oscillation
frequency), and $\lambda$ acting as the square of the gap (i.e.\
pairing field).  The solution of the equation for $a$ gives a train of
hyperbolic secant pulses with a period $T_{\text{pulse}}=2
\ln(\sqrt{\lambda})/\sqrt{\lambda}$; such pulses can be seen in the
time dependence of physical observables such as the occupation of the
bosonic mode $\langle n_{phot}\rangle = a^2$.  For the special case of
$\epsilon_i=0$, Eq.~(\ref{eq:1}) has the solution $\mu=0$ and
$\lambda=N$.

\begin{figure}[htpb]
  \centering
  \includegraphics[width=3.2in]{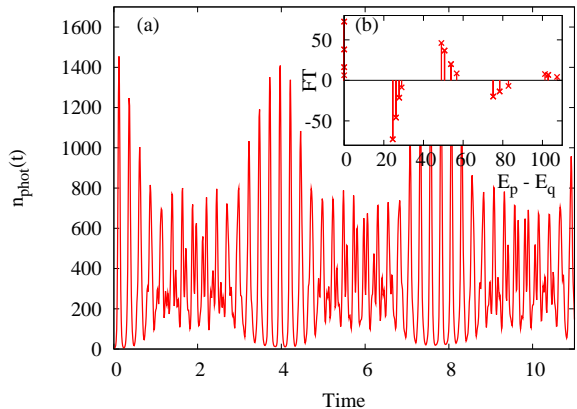}
  \caption{(Color online) Quantum dynamics of the number of photons
    for 2000 spins. Panel (a): time dependence of photon
    amplitude. Panel (b) Principal Fourier components of the time
    dependence.  The lowest seven eigenvalues are sufficient to
    describe the time dependence to better than visible resolution on
    this scale.  All energies and times are in units of the coupling
    between two-level systems and the bosonic mode.}
  \label{fig:simple-trace}
\end{figure}

The problem this paper addresses can be seen most clearly in
Fig.~\ref{fig:simple-trace}, which shows the dynamics of the
population of the bosonic mode $n_{phot}$ according to
Eq.~(\ref{eq:dicke}) with $\epsilon_i = 0$.  The semiclassical train
of hyperbolic pulses is seen in Fig.~\ref{fig:simple-trace}, but there
is in addition a slow envelope, not predicted by the semiclassical
equations, and it is this slow envelope which is discussed below.

By writing the photon number as a sum over eigenstates:
\begin{equation}
  \label{eq:8}
  n(t) = \sum_{pq} 
  \langle 0|X_p \rangle
  \langle X_p|\hat{n}|X_q \rangle
  \langle X_q|0 \rangle
  e^{i(E_p - E_q)t},
  \end{equation}
one may note that the semiclassical result, with its perfectly periodic
train of pulses, corresponds (for $\epsilon=0$) to:
\begin{equation}
  \label{eq:21}
  E_q = q \Omega, \quad \Omega = \pi \frac{\sqrt{N}}{\ln(\sqrt{N})}.
\end{equation}
The inset of Fig.~\ref{fig:simple-trace} shows the Fourier transform
of $n_{phot}(t)$. In contrast to Eq.~(\ref{eq:21}), the eigenvalues of
the full quantum problem are not equally spaced, and so the sidebands
seen in Fig.~\ref{fig:simple-trace}(b) arise.  The period of the slow
envelope is given by the splitting of these sidebands.  Hence, to
describe the approach to the semiclassical result, one must consider
how the deviation from regular spacing $E_q = q \Omega$ (i.e.\ the
anharmonicity of the exact eigenvalues) collapses as $N \to \infty$.

Describing the quantum dynamics by finding the eigenstates is
considerably simplified, since the initial state overlaps with only a
small number of eigenstates.  As discussed in Ref.~\cite{faribault},
the initial condition chosen here satisfies $\langle 0 | H | 0 \rangle
= 0, \langle 0 | H^2 | 0 \rangle= N$, while the semiclassical
eigenstates have $E_q = q \pi \sqrt{N}/\ln(\sqrt{N})$, which implies
that only $\mathcal{O}(\ln(\sqrt{N})^2)$ eigenstates near $|E_q| = 0$
can have any significant overlap with the initial state.  The validity
of this is confirmed by noting that the result of exact
diagonalisation and the result restricting the summation in
Eq.~(\ref{eq:8}) to the seven smallest values of $|E_q|$ are
indistinguishable by eye on the scale of Fig.~\ref{fig:simple-trace}.

\section{WKB approximation and scaling of correction}
\label{sec:wkb-appr-scal}

To find the eigenvalues of the quantum problem, one may approach the
problem by a method closely related to that of \citet{bonifacio70}. As
discussed above, this paper considers the symmetric case, $\epsilon_i
= \epsilon$, for which the Hamiltonian becomes $H=\epsilon S^z + S^- a
+ S^+ a^\dagger$, where $\vec{S} = \sum_i \vec{s}_i$, and as a result,
the quantum state may be written as a wavefunction in the one
dimensional space of occupation, as used in Ref.~\cite{bonifacio70} to
show how the semiclassical limit can arise.  To find not only the
semiclassical limit, but also the corrections to it, one may solve
this one-dimensional problem using a discrete WKB
approximation~\cite{landauQM}.  Using the basis $|n\rangle
= |n_{phot} = n, |\vec{S}|=N/2, S_z = N/2 - n\rangle$, the equation
$(E-\epsilon N/2) \Psi= H \Psi$ becomes:
\begin{equation}
  \label{eq:3}
  (E  + \epsilon n )
  \psi_n = 
  n \sqrt{N+1 - n} \psi_{n-1} + (n+1) \sqrt{N - n} \psi_{n+1}
\end{equation}
The WKB approach consists of two parts; finding the WKB form of the
wavefunction for $(n, N-n) \gg 1$, and then matching this wavefunction
to appropriate forms for $n \simeq 0$ and $n\simeq N$.  In the
following, this matching is referred to as matching the ``boundary
conditions'' for the wavefunction at $n=0, n=N$, but these boundary
conditions are just the Schr\"odinger equation in Eq.~(\ref{eq:3}),
evaluated at $n=0, n=N$, for which those approximations valid for $(n,
N-n) \gg 1$ no longer hold.  This section will consider the case
$\epsilon=0$, the effect of non-zero $\epsilon$ is discussed in
section~\ref{sec:detuning}.

The WKB form of the wavefunction can be derived by considering a
transformation $\psi_n \to \tilde{\psi}_n (i)^n$ which gives the right
hand side of Eq.~(\ref{eq:3}) as a discrete derivative with a variable
prefactor, i.e.\ $E \tilde{\psi} = -i v(x) \partial_x
\tilde{\psi}(x)$.  Written this way suggests a WKB wavefunction of the
form~\cite{landauQM} $\tilde{\psi} \approx [1/\sqrt{v(x)}] \exp[i \int
dz E/v(z)]$, or the discrete analog of this wavefunction.

To write the wavefunction compactly, it is useful to introduce 
the function defined in Ref.~\cite{bonifacio70}:
\begin{equation}
  \label{eq:22}
  g_k \equiv  \frac{\sqrt{2}\Gamma(1+k/2)}{\Gamma[1/2+k/2]}
  \approx
  \begin{cases}
    \sqrt{2/\pi} & k = 0 \\
    \sqrt{\pi/2} & k = 1 \\
    \sqrt{k+1/2} & k \gg 1 
  \end{cases}
\end{equation}
which is chosen such that $g_k g_{k+1} = k+1$.  With this notation,
the WKB wavefunction can be written:
\begin{equation}
  \label{eq:4}
  \psi_n^{(WKB)} = \frac{%
    \cos\left(E \Phi_n + \phi + n \pi/2 \right)
  }{\sqrt{g_n^2 g_{N-n}}}, 
\end{equation}
where 
\begin{math}
  \Phi_N = \sum_{m=n}^N 1/2 g_m^2 g_{N-m},
\end{math}
and $\phi$ is an arbitrary phase set by boundary conditions.  This
wavefunction is valid while the change of phase between two successive
values of $n$ is much less than one; this condition for the validity
of the WKB wavefunction can formally be written as $g_n^2 g_{N-n} \gg
E$.  Under the same conditions, one may approximate the summation in
the definition of $\Phi_n$ by integration, and hence to leading order
one has:
\begin{equation}
  \label{eq:5}
  \Phi_n 
  \approx
  \frac{1}{\sqrt{N+1}} \Acosh\left[
    \sqrt{\frac{N+1}{n+\frac{1}{2}}} \right].
\end{equation}
It is worth noting that for even $N$, the solution in Eq.~(\ref{eq:4})
with $E=0$ can be seen to be an exact eigenstate of Eq.~(\ref{eq:3}).

\subsection{Matching WKB solution at boundaries}
\label{sec:match-wkb-solut}

\subsubsection{Boundary condition at $n=N$}
\label{sec:bound-cond-at}

When considering the boundary conditions, the boundaries at $n=0$ and
$n=N$ behave differently.  At $n=N$  first note that $g_N^2 g_{0}
\simeq (N+1/2) \sqrt{2/\pi} \gg E \simeq \pi \sqrt{N}/\ln(\sqrt{N})$,
meaning the WKB wavefunction is valid right up to this boundary.  In
addition, the Schr\"odinger equation at the boundary becomes $E \psi_N
= N \psi_{N-1}$, and so to order $1/\sqrt{N}$, this boundary condition
is satisfied by $\psi_{N-1} = 0$.  Since the corrections to the
semiclassical energies found below are of order $1/\ln(\sqrt{N})$, the
much smaller corrections of order $1/\sqrt{N}$ can be neglected, and
so the boundary condition at $n=N$ requires
\begin{math}
  \phi = (2q+2-N)\pi/2
\end{math}
where $q$ is any integer; for even $N$ one may thus choose $\phi=0$.
From herein, even $N$ is assumed; for odd $N$ similar results with
$\phi = \pi/2$ follow straightforwardly.

\subsubsection{Boundary condition at $n=0$}
\label{sec:bound-cond-at-0}

The boundary condition at $n=0$ is more complicated.  For this
boundary, $g_0^2 g_{N} \simeq (2/\pi) \sqrt{N+1/2}$, and so the
condition $g_0^2 g_{N} \gg E \simeq \pi \sqrt{N}/\ln(\sqrt{N})$ is not
necessarily satisfied --- the validity of the WKB wavefunction depends
on taking $\ln(\sqrt{N}) \gg 1$, which requires very large $N$; these
corrections due to finite $\ln(\sqrt{N})$ are the quantum corrections
of interest in this paper.  

Because the WKB wavefunction breaks down for small values of $n$, it
is necessary to match the WKB wavefunction to the exact wavefunction
at a non-zero value of $n=n_0$, rather than at $n=0$; i.e the exact
wavefunction should be calculated for $n\le n_0$, and the WKB
wavefunction made to match it at $n=n_0$.  The larger the value of
$n_0$ one takes, the better the calculated energies will match the
exact solution.  Surprisingly, even matching the solution at $n=0$
turns out to provide useful information, and correctly reproduces the
scaling of energy with system size; this is discussed below in
Sec.~\ref{sec:simpl-bound-scal}.

\begin{figure}[htpb]
  \centering
  \includegraphics[width=3.2in]{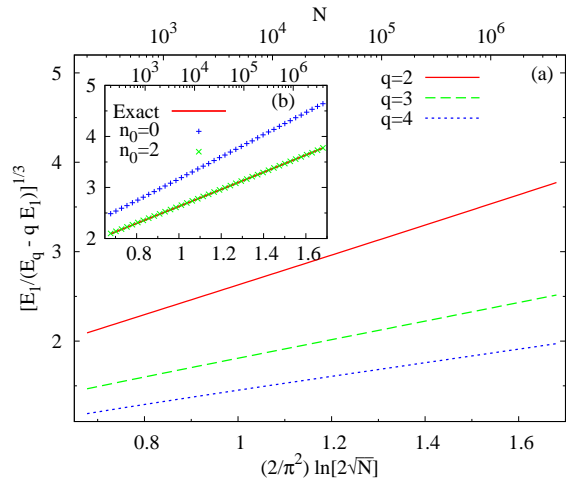}
  \caption{(Color online) Anharmonicity, $E_q - q E_q$, plotted
    against system size $N$ with variables transformed to demonstrate
    agreement with the scaling expected from Eq.~(\ref{eq:7}).  Panel
    (a) compares $E_q$ for $q=2,3,4$. Panel (b) compares WKB
    approximations to $E_2 - 2 E_1$, on the same transformed axes,
    with $n_0$ terms near the $n=0$ boundary.}
  \label{fig:scaling}
\end{figure}

The procedure of matching is straightforward; for a given value $n_0$,
one finds the exact solution $\psi^{(0)}_{n\le n_0}(E)$ by solving
Eq.~(\ref{eq:3}), with $\psi_{n_0+1} = \psi_{n_0+1}^{(WKB)}$; i.e.\:
\begin{equation}
  \label{eq:10}
  E \psi^{(0)}_n = H^{(n_0)}_{nm} \psi^{(0)}_m
  + \delta_{n,n_0} (n_0+1) \sqrt{N-n_0} \psi_{n_0+1}^{(WKB)},
\end{equation}
where $H^{(n_0)}$ is the Hamiltonian restricted to $n\le n_0$; the
matching condition then becomes $\psi^{(0)}_{n_0} =
\psi_{n_0}^{(WKB)}$.  By diagonalising the $(n_0 +1) \times (n_0 + 1)$
matrix $H^{(n_0)}$ to give $H^{(n_0)} = \lambda_{\alpha} |
\chi_{\alpha} \rangle \langle \chi_{\alpha} |$, the boundary condition
equation can be written:
\begin{multline}
  \label{eq:11}
  \cos\left( \frac{n_0 \pi}{2} + \Phi_{n_0} \right)
  =
  -
  \sin\left( \frac{n_0 \pi}{2} + \Phi_{n_0+1} \right)
  \\
  \times
  g_{n_0}^2 g_{N-n_0}
  \sum_{\alpha} 
  \frac{|\langle\chi_{\alpha}|n_0\rangle|^2}{E-\lambda_{\alpha}}.
\end{multline}
This gives a nonlinear equation for $E$, the complexity of which
increases with increasing $n_0$.  An example of solving this equation,
with $n_0=2$ is shown in Fig.~\ref{fig:scaling}(b), which clearly
accurately reproduces the results of exact diagonalisation.

\subsection{Simplified boundary condition and scaling with system size}
\label{sec:simpl-bound-scal}

To understand the how the corrections to $E_q$ depend on number of two
level systems $N$, the equation with general $n_0$ is rather
complicated, but as seen in Fig.~\ref{fig:scaling}(b), matching at
$n=0$ gives the correct scaling with system size, but an incorrect
numerical coefficient. Considering this matching at $n=0$, the
boundary condition becomes from Eq.~(\ref{eq:3}),~(\ref{eq:4}):
\begin{equation}
  \label{eq:6}
  E \cos( E \Phi_0) = - \frac{2}{\pi} \sqrt{N} \sin(E \Phi_1).
\end{equation}
In Fig.~\ref{fig:scaling}(b), the solution of this equation is represented
by the blue crosses, labelled $n_0=0$.

To the same level of approximation made so far, one may set $\Phi_0
\simeq \Phi_1 \equiv \ln(\sqrt{N}) /\sqrt{N}$.  If $\ln(\sqrt{N})$ were
large, the solution of Eq.~(\ref{eq:6}) would be $E_q = q \pi/\Phi_1 
= q \pi
\sqrt{N}/ \ln(\sqrt{N})$, i.e.\ the semiclassical result.  For finite
$\ln(\sqrt{N})$, Eq.~(\ref{eq:6}) produces corrections
$\mathcal{O}(q^3)$ that break the harmonicity; these terms look like:
\begin{equation}
  \label{eq:7}
  \delta E_q = C q^3 \frac{\sqrt{N}}{[\ln(\sqrt{N})]^4}.
\end{equation}
The coefficient $C$ as calculated by expanding Eq.~(\ref{eq:6}) in the
small parameter $1/\ln(\sqrt{N})$ is $C=\pi^6/24 \simeq 40.1$ while
the best fit coefficient over the range shown in
Fig.~\ref{fig:scaling}(a) is $C=13.5$, howevever this gradient may
should be treated with caution, as $1/\ln(\sqrt{10^6})$ is not a small
number.  The form in Eq.~(\ref{eq:7}) is the first main result of this
paper --- as shown in Fig.~\ref{fig:scaling}, this dependence on
$\ln(\sqrt{N})$ well describes the scaling of the anharmonicity;
because the dependence on $N$ is so slow, quantum corrections to the
semiclassical result remain relevant even for $10^6$ two-level
systems.

\section{Detuning}
\label{sec:detuning}

\begin{figure}[htpb]
  \centering
  \includegraphics[width=3.2in]{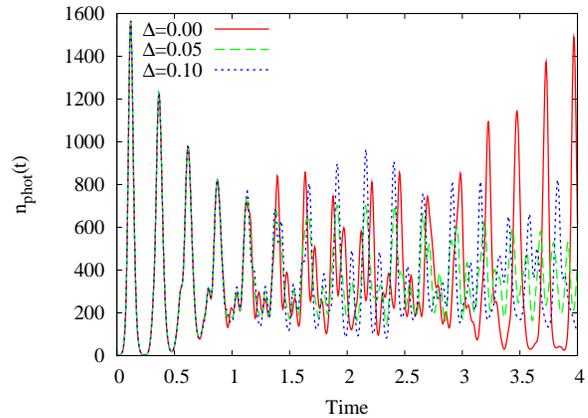}
  \caption{(Color online) Time dependence of photon number for
    differing values of detuning $\epsilon$, showing modification to
    the slow envelope.}
  \label{fig:detuning-trace}
\end{figure}

The remainder of this paper discusses the effect of detuning, i.e.\  of
$\epsilon_i = \epsilon \ne 0$; this again reveals a distinction
between the semiclassical dynamics and the full quantum mechanical
problem, and provides some insight into the case where not all values
of $\epsilon_i$ are identical.  The time dependence in the case of
$\epsilon_i = \epsilon \ne 0$ is shown in
Fig.~\ref{fig:detuning-trace}, showing a change to the slow envelope.
Semiclassically, the effect of non-zero $\epsilon$ in Eq.~(\ref{eq:1})
is to give $\mu=\epsilon/2$, $\lambda = N - (\epsilon/2)^2$. Since it
is $\lambda$ that controls the collective oscillations, the leading
correction to the frequency of pulses, i.e.\ the semiclassical energy
splitting $\Omega$ should be quadratic in $\epsilon$:
\begin{equation}
  \label{eq:20}
  E_q = q \Omega,\quad
  \delta \Omega
  \simeq \frac{\pi \epsilon^2 }{8 \sqrt{N} \ln(\sqrt{N})}.
\end{equation}
The following discussion focusses on the case of small $\epsilon$, and
shows that quantum corrections give shifts of energies linear
in $\epsilon$ for small $\epsilon$, such that for any finite $N$, the
leading correction to $E_q$ is quantum, not semiclassical.

It is straightforward to see from Eq.~(\ref{eq:3}) that for small
$\epsilon$, the linear peturbative correction to eigenstate $q$ would
be:
\begin{equation}
  \label{eq:9}
  E_q \to E_q - \epsilon \langle X_q | \hat{n} | X_q \rangle,
\end{equation}
and so the aim of this section is to find how $\langle n_q \rangle
\equiv \langle X_q | \hat{n} | X_q \rangle$ depends on system size $N$
and eigenstate label $q$.  One should note that a q independent shift
of $E_q$ has no effect on the time dependence of photon number.  In
the semiclassical picture, such a constant shift just corresponds to a
change of $\mu$, hence our interest is in the quantity $\delta E_q -
\delta E_0$.

\begin{figure}[htpb]
  \centering
  \includegraphics[width=3.2in]{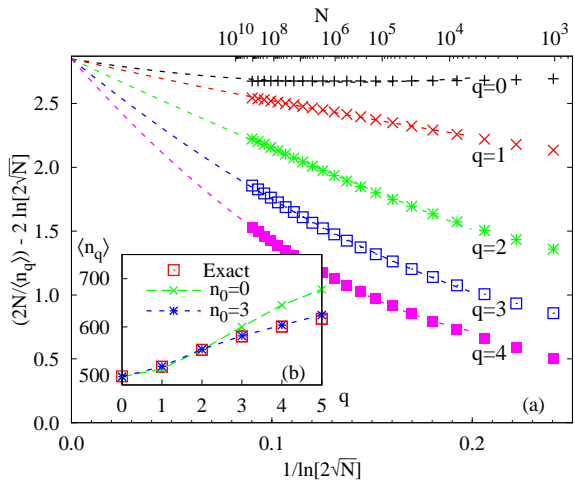}
  \caption{(Color online) Panel (a): Scaling of $\langle n_q \rangle$
    with system size calculated using the WKB wavefunction matched to
    the exact solution for $n<n_0=40$, and transformed following
    Eq.~(\ref{eq:18}). Dotted lines are a quadratic best fit to the
    data for $N>10^4$, assuming all lines tend to a common asymptote.
    Panel (b): comparison of exact diagonalisation and WKB
    approximations for $\langle n_q \rangle$ for $N=3000$, showing
    that the WKB approximation should be valid for the panel (a).}
  \label{fig:detuning-scaling}
\end{figure}

Figure~\ref{fig:detuning-scaling}(b) shows that the WKB eigenstates
matched to the exact solution for $n<n_0=3$ reproduce the result of
exact diagonalisation for $N=3000$, giving confidence that the WKB
wavefunction can be used to determine $\langle n_q \rangle$.  Using
the WKB wavefunction of Eq.~(\ref{eq:4}) one has:
\begin{equation}
  \label{eq:12}
  \langle n_q \rangle
  \simeq
  \frac{%
    \displaystyle
    \sum_{n=0}^N n \frac{%
      \left[ 1 + (-1)^n \cos(2 E_q \Phi_n) \right]
    }{g_n^2 g_{N-n}}
  }{%
    \displaystyle
    \sum_{n=0}^N \frac{%
      \left[ 1 + (-1)^n \cos(2 E_q \Phi_n) \right]
    }{g_n^2 g_{N-n}}
  }.
\end{equation}

To analyse how the above expression depends on the number of two-level
systems $N$, it is convenient to write it as $\langle n_q \rangle =
(A_0 + \delta A_q) / (B_0 + \delta B_q)$, where $\delta A_q, \delta
B_q$ come from the term proportional to $(-1)^n$ in the summand, and
$A_0$, $B_0$ from the other term.  One may then show that:
\begin{equation}
  \label{eq:13}
  A_0 = \sum_{n=0}^N \frac{n}{g_n^2 g_{N-n}}
  \approx
  \int_0^n dn \frac{1}{\sqrt{N-n}} = 2 \sqrt{N}
\end{equation}
and from Eq.~(\ref{eq:5}), one sees that 
\begin{equation}
  \label{eq:14}
  B_0
  \approx \frac{2}{\sqrt{N}} \Acosh(\sqrt{2N})
  \approx \frac{2}{\sqrt{N}} \ln(2\sqrt{2N}).
\end{equation}
Because $A_0 \gg B_0$, one may anticipate that $\delta B_q$ has a more
significant effect on the answer than does $\delta A_q$; this can be
made more firm by noting that for even $N$:
\begin{displaymath}
  \label{eq:15}
  \delta A_{q=0} \approx
  \sum_{n=0}^N \frac{(-1)^n}{\sqrt{N-n+1/2}}
  \approx
  \sum_{n=0}^{\infty} \frac{(-1)^n}{\sqrt{n+1/2}} \approx 0.944,
\end{displaymath}
and numerically one may confirm that $\delta A_{q}$ is a factor
$\sqrt{N}$ smaller than $A_0$ in general.  For $\delta B_{q}$ one may
observe that the main contribution to the sum comes from values of $n
\ll N$, where the denominator is smallest, and so approximate:
\begin{equation}
  \label{eq:16}
  \delta B_q \approx 
  \frac{1}{\sqrt{N}}
  \sum_{n=0}^{\infty} \frac{%
    (-1)^n \cos(E_q \Phi_n) 
  }{(n+1/2)}.
\end{equation}
This expression for $\delta B_q$ has the form of $1/\sqrt{N}$ multiplied
by a function that should depend only on $\ln(\sqrt{N})$ --- by considering
the forms of $E$ in Eq.~(\ref{eq:21}) and $\Phi_n$ in Eq.~(\ref{eq:5}), it
is clear that only logarithmic dependence on system size enters the
product $E \Phi_n$.

 Putting all these considerations together one may write:
\begin{displaymath}
  \label{eq:17}
  \langle n_q \rangle \approx 
  \frac{2 N}{2 \ln(\sqrt{N}) + \sqrt{N} \delta B_q}
\end{displaymath}
and thus one may expand:
\begin{align}
  \label{eq:18}
  \frac{2N}{\langle n_q \rangle} - 2 \ln(\sqrt{N}) 
  &\approx \sqrt{N} \delta B_q \nonumber\\
  &=
  \beta_q^{0} 
  + \frac{\beta_q^{1}}{\ln(\sqrt{N})}
  +  \frac{\beta_q^{2}}{\ln(\sqrt{N})^2}
  + \ldots
\end{align}
for large $N$.  This scaling is indeed seem to occur in
Fig.~\ref{fig:detuning-scaling}(a), and strongly suggests that
$\beta_{q}^0$ is independent of $q$, while $\beta_{q}^1$ depends on
$q$.  One may thus write the $q$ dependent part of the energy shift
as:
\begin{equation}
  \label{eq:19}
  \delta E_q - \delta E_0
  \approx
  -  \frac{\epsilon N}{\ln(\sqrt{N})}
  \left(
    \frac{\beta^1_0 - \beta^1_q}{2 \ln(\sqrt{N})^2}
  \right) + \ldots,
\end{equation}
where only the leading order term in powers of the logarithm have been
retained.  This expression should be compared to the classical
correction in Eq.~(\ref{eq:20}).

Comparing Eq.~(\ref{eq:19}) to the semiclassical correction of
Eq.~(\ref{eq:20}), one may note that in both the limit fixed $N$,
$\epsilon \to 0$ and also for fixed $\epsilon$, $N \to \infty$, the
quantum correction is larger than the semiclassical correction.

\subsection{Relating detuning to disordered $\epsilon_i$.}
\label{sec:rela-detun-disord}

The above results on treating $\epsilon$ perturbatively can also
describe properties of the model where $\epsilon_{i=1 \ldots N-1} =
0$, $\epsilon_N=\epsilon$; i.e.\ 
\begin{math}
  \delta H = \epsilon s_N^z,
\end{math}
again considering $\epsilon$ perturbatively.  The fact this case can
be treated by first order non-degenerate perturbation theory is not
trivial; it arises because although degenerate states do exist
(particularly states with zero energy), the asymmetry introduced by
altering a single spin energy $\epsilon_i$ does not mix these
degenerate states.  With a greater number of different $\epsilon_i$
this simplifying condition fails, and degenerate perturbation theory
is instead required.

For the special form of $\delta H$ above for which non-degenerate
perturbation theory is relevant, one may consider the the expectation
of perturbation Hamiltonian between the boson number states, for which
one may show that
\begin{equation}
  \label{eq:23}
  \langle n | s_i^z | n^{\prime} \rangle = \delta_{nn^\prime} 
  \left( \frac{1}{2} - \frac{n}{N} \right)
\end{equation}
which reproduces the results of Sec.~\ref{sec:detuning} with $\epsilon
\to \epsilon/N$.  Such a result hints that quantum corrections in the
non-symmetric model may become more significant, since for weak
disorder of energies $\epsilon_i$, such linear corrections due to
quantum corrections will win over the quadratic semiclassical effects
of disorder.  However, direct calculation of the quantum dynamics is
challenging, as the size of the Hilbert space explored in such a model
is exponential in $N$, and if quantum corrections still vanish as
powers of $\ln(\sqrt{N})$, one requires very large systems to study
the asymptotic behaviour at large $N$.

\section{Conclusions}
\label{sec:conclusions}

In conclusion, the quantum dynamics of the symmetric Tavis-Cummings model
starting from an initially inverted state describes a train of
hyberbolic secant pulses (as in the semiclassical result), but with an
additional slow envelope.  The slow envelope arises due to
the anharmonicity of the eigenvalues of the quantum problem; this
anharmonicity reduces as the system size increases, but only
logarithmically with the number of two level systems.  In the absence
of detuning, i.e.\ for $\epsilon_i=0$, the eigenvalues take the form
$E_q \simeq q \pi \sqrt{N}/\ln(\sqrt{N}) + C q^3
{\sqrt{N}}/{[\ln(\sqrt{N})]^4}$.

With a small detuning $\epsilon$, the quantum problem has a
perturbative correction linear in the size of detuning $\delta E
\propto \epsilon N / [\ln(\sqrt{N})]^3$ , while the semiclassical
result has only corrections quadratic in the detuning $\delta E
\propto \epsilon^2 / [\sqrt{N} \ln(\sqrt{N})]$.  As a result, for
either finite detuning and $N\to \infty$ or finite $N$ and $\epsilon
\to 0$, the classical effects vanish faster than quantum corrections.
This result may also be indicative of the effects of disorder, i.e.\
of non-identical values of $\epsilon_i$.  For a fixed distribution of
$\epsilon_i$, as $N\to \infty$ the effects of this disorder vanish
compared to the energy scale of the common coupling $\sim
\sqrt{N}$. It seems likely that in such a limit, quantum corrections
may vanish more slowly than the classical effects of the distribution
of $\epsilon_i$.

In the general case of disordered $\epsilon_i$, the quantum problem is
significantly harder to solve, as the size of the Hilbert space grows
exponentially with the number of two level systems, whereas for
$\epsilon_i=\epsilon$, it grows only linearly.  However, the
integrability of the problem may simplify the problem, as discussed by
\citet{faribault}.  In the symmetric case, the integrability of the
semiclassical problem means that while conservation laws restrict the
system to exploring an $N$ dimensional Hilbert space, in the limit of
large $N$, the system in fact only explores a one dimensional path
through this space.  Similarly for the disordered model, the problem is
integrable~\cite{dukelsky04,kundu04}, which in turn leads to the
semiclassical dynamics~\cite{andreev04,barankov04,yuzbashyan05:prb}
following a one dimensional path, suggesting that a simple description
of the quantum corrections even for large values of $N$ might be
possible.

Note added: Since the submission of this work, a similar treatment of
this model has been undertaken by \citet{babelon09}.

\begin{acknowledgments}
  I would like to acknowledge useful discussions with P.~R.~Eastham,
  M.~J.~Bhaseen, J.~Hope  and to acknowledge funding under
  EPSRC grant no EP/G004714/1.
\end{acknowledgments}


\end{document}